\documentclass[10pt,letterpaper]{book}
\usepackage{frontier07}
\begin{document}
\newcounter{ctr}
\setcounter{ctr}{\thepage}
\addtocounter{ctr}{8}

\talktitle{Why Tau First?}
\talkauthors{Daniele Fargion \structure{a,b}, Daniele D'Armiento \structure{a}
             Pier Giorgio Lucentini De Sanctis \structure{a}.}

\begin{center}
\authorstucture[a]{Dipartimento di Fisica,
                   Universit\`a di Roma "Sapienza",  \newline
                   Pzle.A.Moro 2, Roma, Italy}

\authorstucture[b]{INFN, Sezione di Roma I,
                    Pzle.A.Moro 2, Roma, Italy}
\end{center}

\shorttitle{Why Tau First?}

\firstauthor{D.Fargion}

\begin{abstract}
Electron neutrino $\nu_e$ has been the first neutral lepton to be foreseen and discovered last century. The un-ordered muon ${\mu}$ and its neutrino $\nu_{\mu}$ arose later by cosmic rays.
 The  tau $\tau$ discover, the heaviest, the most unstable charged lepton,
 was found surprisingly on ($1975$). Its $\nu_{\tau}$ neutrino was hardly revealed just on ($2000$). So why  High Energy Neutrino Astronomy should rise first via $\nu_{\tau}$, the last, the most rare one?  The reasons are based on a chain of three favorable coincidences found last decade: the  neutrino masses and their flavor mixing, the UHECR opacity on Cosmic Black Body
 (GZK cut off on BBR), the amplified  $\tau$ air-shower decaying in flight.
 Indeed guaranteed UHE GZK ${{\nu}_{\tau}}$, ${\overline{{\nu}_{\tau}}}$
  neutrinos, feed by muon mixing, while skimming the Earth might lead to boosted
  UHE $\tau$,${\overline{{\tau}}}$, mostly horizontal ones. These UHE $\tau$
  decay in flight are spread, amplified, noise free Air-Shower: a huge event for an unique particle.   To be observed soon: within Auger sky, in present decade. Its discover may sign of the first tau appearance.
\end{abstract}
\section{The Cosmic  multi-frequency spectra  up to GZK edges}
\begin{figure}
\begin{center}
\epsfig{figure=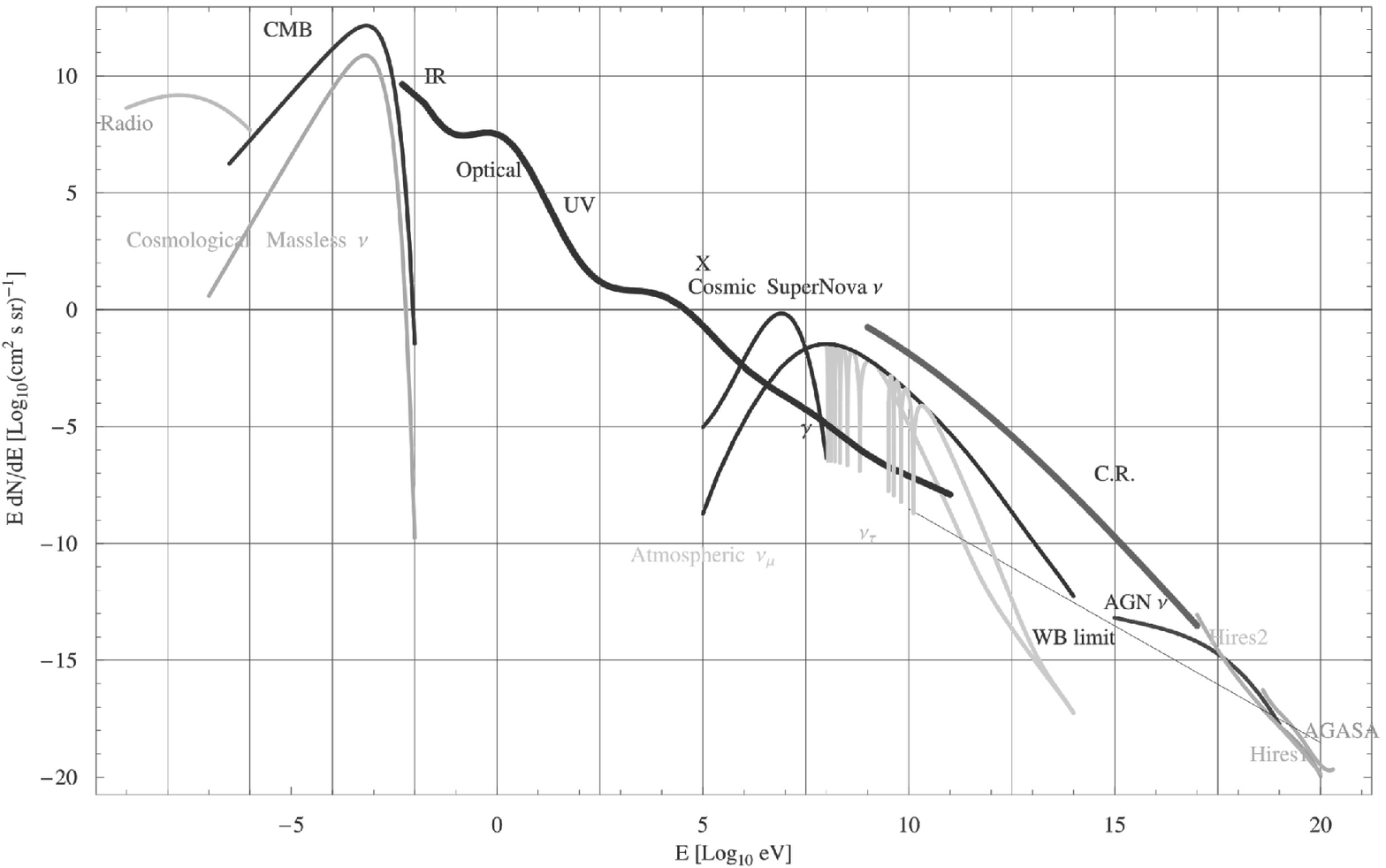,width=8.5cm}
\end{center}
\caption{The flux number multi-frequency panorama of cosmic radiations.
 Solar and local galactic components $\gamma,\nu$ have been omitted.
 Both  atmospheric $\nu_{\mu}$ and its parasite oscillated $\nu_{\tau}$ component are shown;
 the twin $\nu_{\tau}$ curves are showing both the vertical, crossing the Earth, (the  one with a deep at $10$ GeV) and the horizontal components in all energy band. }
   \label{Fig1}
\begin{center}
\epsfig{figure=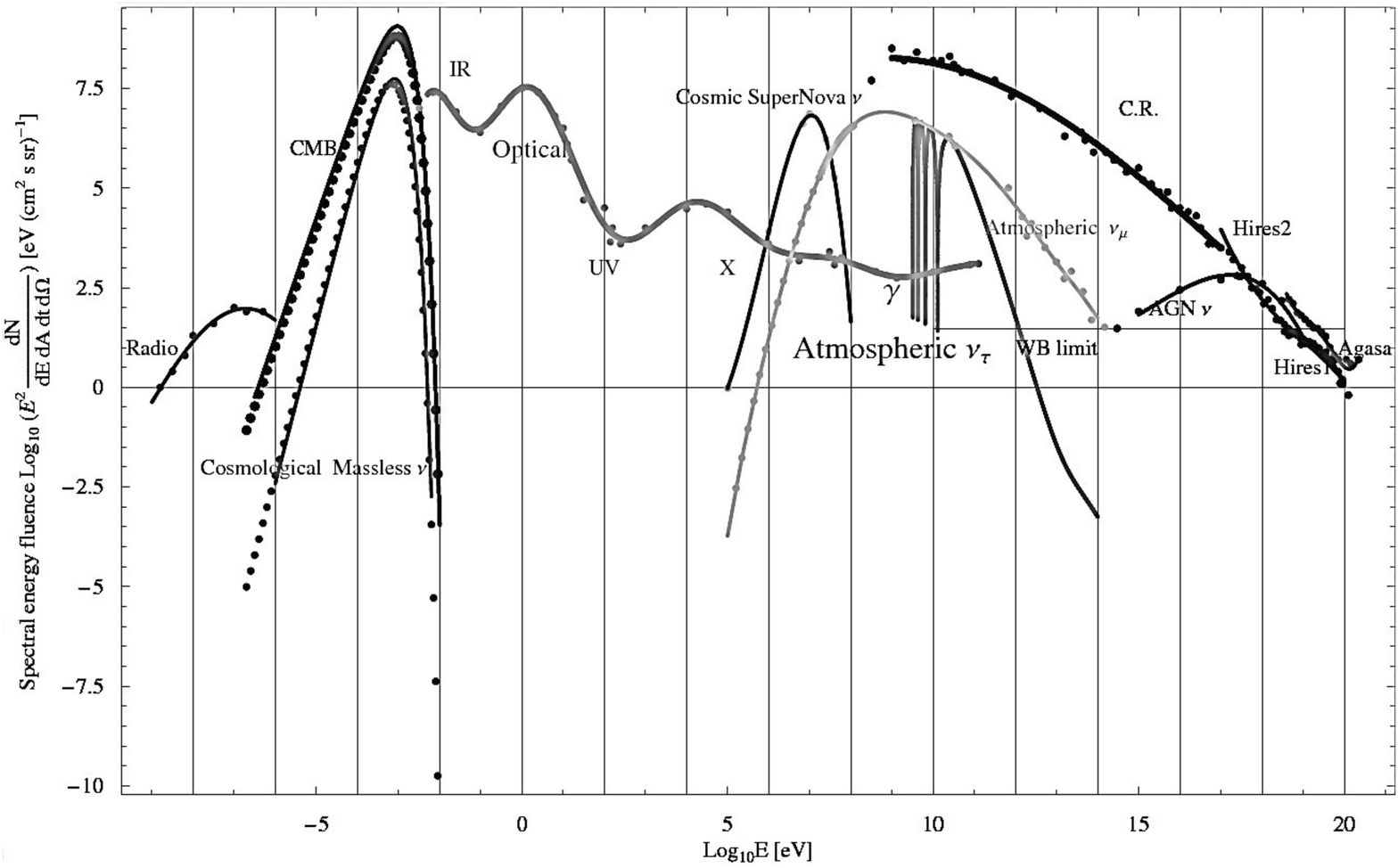,width=9.5cm}
\end{center}
\caption{ The energy spectra  of the cosmic radiation consequent
of the number flux in previous figure; theoretical and few
observed data points are shown. Only the vertical $\nu_{\tau}$ are
shown. Above TeV they are free of atmospheric noise. The GZK cut
off at the extreme is the source of the GZK neutrino, at WB range,
discussed in the article. }
   \label{Fig2}
\end{figure}

    High energy neutrino astronomy at GZK \cite{Greisen:1966jv},\cite{za66} limit is ready to be discovered. Its role may shine light in Universe understanding.  Our present view of this  Universe is summarized in radiation flux number spectra updated in Fig.\ref{Fig1}; its consequent energy fluency spectra  in considered in Fig.\ref{Fig2}.   The flux number spectra ranges from radio frequency to cosmic Black Body Radiation, BBR,  toward Cosmic Rays up to Ultra High Energy ones, UHECR. For calibration also a BBR Fermi-Dirac  for any eventual massless cosmic neutrino (at $1.9 K$ temperature). Light neutrino with mass  are probably non-relativistic and  gravitationally clustered, not displayed here. The role in cosmology of relic neutrinos has been widely reviewed recently \cite{Dolgov2002}.  Infrared and the Optical photons fluxes are followed by UV and X cosmic ones.  Last $\gamma$ Astronomy is at MeV-GeV-TeV band edge. In MeV region,
      Supernova Cosmic  relic Neutrinos background  may soon arise. Unfortunately the CR  secondaries,  $\pi^{\pm}$,$\mu^{\pm}$  are blurred as well as their $\nu_{\mu}$ and $\nu_e$
       final atmospheric neutrinos: this cause to $\nu_{\mu}$ and $\nu_e$ astronomy to be also smeared and polluted.
    The inclined line  on edge Fig. \ref{Fig1} tag the so-called Waxmann- Bachall limit,$\simeq E^{-2}$ as  a minimal   limit for GRB and GZK  $\nu$ flux  expectation. It is easy to note as
     this fluency is correlated to average cosmic radio background, as well as
     it is comparable to UHECR at ten EeV band and average GRB fluency.
     There are good reasons to foresee a WB GZK neutrino    background too.   The coexistence on many cosmic radiations makes the known windows on the     Universe an exciting growing puzzle. Neutrino astronomy at different band   may offer the key answers.  Indeed while photons are  neutral, un-deflected,  offering Astronomy pictures,  most of the Cosmic Ray are charged and smeared by galactic  and cosmic magnetic fields.  Therefore Cosmic Rays (CR) offer only an integrated, short-sighted Astrophysics.  The presence of galactic magnetic fields are  reminding us of the puzzling  absence of magnetic monopoles in our Universe.
           The low energy multi-frequency spectra on left  side (below TeV energy)
            is dominated by photons; at higher energies  (TeV-PeVs) the photons are rarer and opaque to relic extra-galactic Infrared photons.  Tens TeV photons arrive only from nearby Universe (hundred Mpc radius); at PeV energy,   the  cosmic Black Body Radiation (BBR) makes UHE photons bounded in our Local Group,  (Mpc) size volume. Therefore only PeVs neutrinos may reach us  from  Universe edges.
     The right side Fig.\ref{Fig1}, the high energy one,  is dominated by Cosmic Rays and its secondaries.   The ruling dominance of solar photons and  of its neutrinos is obviously hidden here  to avoid confusions.Indeed beyond the MeV energies, where solar neutrino flux dominates, one expects a peculiar  niche for the relic Supernova background, still on the edge of detection.  At tens MeV atmospheric neutrino noise will pollute this SN signal.
    Hopefully upgraded underground SK (Super Kamiokande) detector might soon reveal the SN trace.
     At the same $10^7$ eV energy band,  very  rare and bright  galactic supernova neutrino,
      as the famous SN1987A,  might rarely blaze our Milky Way almost once a century.
    Future Megaton Neutrino detectors could observe even nearby Andromeda Supernova,
    making three times larger the previous rate.  In the same energy range, much less power-full
     but more frequent neutrino burst, in all flavor,  may rise  from largest solar flares,
      once in a decade. They are  better detectable  by  noise free anti-neutrino electron $\overline{\nu}_e$  component in Megaton detectors. Among  the cosmic rays, the secondary Atmospheric Neutrinos arise, blurred and noisy as their parental CR. Therefore Neutrino muon and electron signals are largely polluted by abundant atmospheric secondaries.
          The charmed pions (rare parents  of tau) are hardly produced by CR respect to  pion and Kaon ($ < 10^{-5}$) \cite{athar05}.  Energetic atmospheric $\nu_{\mu}$ cannot feed $\nu_{\tau}$  via neutrino mixing, because too short distance for known mass splitting.  Then atmospheric HE $\nu_{\tau}$ are suppressed and its astronomy is  noise free.  Unfortunately  TeVs Tau neutrino are still difficult to be disentangled from other  neutral current neutrino events. But higher energy ones, in PeV-EeV band, may reveal themselves loudly \cite{Fargion1999}. As discussed below.

\section{The Auger-Hires spectra: GZK cut-off, expected  UHE $\nu$   Fluxes}
   Up to day the puzzles on CR and on UHECR  remain unsolved:
   what are the sources, how they are accelerated, is there any GZK cut-off,
    why local UHECR sources are not yet observed?   Agasa-Hires and Auger moved the problem answer randomly  from one edges to another. BL-Lac connection  with UHECR, found first by Agasa \cite{Gorbunov02} and confirmed somehow by Hires in last few years \cite{Hires06}, apparently fade away by Auger null results. Clustering events too . The early Agasa Galactic Anisotropy at EeV, hint for a timid, but relevant, new Galactic Neutron Astronomy,  disappeared under Auger scrutiny.     Moreover a surprising  composition record in Auger UHECR  data is   unexpected:  a turn toward heavy (Fe) nuclei at highest energy events. They may produce less neutrinos,  if they are very local (but than, what are their arrival directions?). Otherwise being isotropic, they are call for a cosmic nature, possibly born at ZeV energy. In this view their photo-nuclear fragility (diffusion distance of few Mpc) imply  once again  a much  abundant UHE GZK neutrino fluxes to be found. The puzzle grows.  The presence of a drastic or at least a mild decrease in UHECR spectra edges arose from Hires and AUGER data. This in contrast with AGASA  hint for the absence of a GZK cut off. The absence of source identifications within a GZK volume pose additional puzzles: are UHECR isotropic and homogeneous (as GRBs), spread along the whole  Universe? How can they overcome the cosmic photon opacity (GZK puzzle)?  To face this possibility
   we \cite{Fargion-Mele-Salis99} did offer a decade ago the Z-Shower or Z-Burst  model \cite{Fargion-Mele-Salis99}. This model  is based on  UHE ZeV neutrinos primary,
  ejected from the cosmic sources as the courier, transparent to BBR photons,
  interacting at the end of the flight, with their relic non relativistic cosmic partners
   clustered in wide cloud as hot dark matter. They are the favorite target of the interaction
   via Z-boson resonance. The UHE Z produced and its decay in flight  would lead to UHE nucleons traces of observed UHECR.  This model got alternate  attention and fortune, but its motivation (the need to overcome isotropic and homogeneity  UHECR spectra)  survived the last Auger test \cite{Yamamoto2007}.  More models able to fit the spectra require a diffuse UHECR protons
       source around ZeV energy \cite{Berezinsky 2002}; such an energy for the primary  in complete agreement  with the Z-shower model versions,  with a tuned neutrino mass at $m_{\nu}= 0.08$ eV, well compatible with cosmological limits and atmospheric mass splitting \cite{Fargion-Mele-Salis99},\cite{Fargion06}.  However, as noted above, last surprising Auger claim of a heavy UHECR composition  is making all these conclusions questionable.
    To estimate a minimal GZK neutrino flux we note that the Auger UHECR at GZK knee
     $ E = 3.98  \cdot 10^{19} eV$ is corresponding to a small fluency
     ($\Phi_{GZK}\simeq 6.6 eV \cdot cm^{-2}s^{-1} sr^{-1}$ ): at its average maxima
     $ E \simeq 1.1  \cdot 10^{20} eV$, the flux is very suppressed  ($\Phi_{GZK}\simeq 0.5 eV \cdot cm^{-2}s^{-1} sr^{-1}$ ); this flux   must suffered  severe losses  along the whole cosmic volumes,  into GZK secondaries,  mostly few EeV GZK neutrinos. A simple estimate may be done based on this flux  amplified by the Universe/GZK size ratio, a value of nearly two order of magnitude.  The final total UHECR GZK fluence estimated in this and other ways is
        ($\Phi_{GZK}\simeq 50 eV \cdot cm^{-2}s^{-1} sr^{-1}$ ), whose main traces are electron pairs,  $\nu$ pairs of all three flavors. This offer, following most authors,
           a neutrino (pair) GZK minimal energy spectra at EeV.
         $\Phi_{\nu_{\tau}+\overline{{\nu_{\tau}}}}\simeq 20 eV \cdot cm^{-2}s^{-1} sr^{-1}$
          to assume for up-going taus. This value may be at worst a half of it, but not too far way. A different, convergent hint of a minimal fluency comes from the UHECR ankle threshold  at $ E = 3.98  \cdot 10^{18} eV$ : it may be mark the crossing from galactic to extragalactic components; it may also mark the electron pair losses;
           it may  also be source of photo-pion production of UHECR  escaping from their bright source. The consequent fluency may exceed ($\Phi_{EeVs}\simeq 25 eV \cdot cm^{-2}s^{-1} sr^{-1}$),compatible with previous fluency value. Therefore for sake of simplicity we assume around EeV energy  a minimal flat ($\propto E^{-2}$) neutrino $\tau$ spectra (the sum of both two species), comparable with the WB one,  at a nominal fluency $\Phi_{\nu_{\tau} +\overline{ \nu_{\tau}}}\simeq 20 eV \cdot cm^{-2}s^{-1} sr^{-1}$.
         For a fluency $50\%$ larger  we derived \cite{Fargion2004}, earlier estimate mainly for EUSO;   we considered in detail the Earth opacity to UHE neutrinos for an
          exact terrestrial density profile: its column depth defined the survival for UHE ${\nu_{\tau}}$ at each zenith angle, the consequent $\tau$ probability to escape   and to decay in flight considering the terrestrial finite size  atmosphere.
          Our result for Auger now, for $\Phi_{\nu_{\tau}+\overline{{\nu_{\tau}}}}\simeq 20 eV \cdot cm^{-2}s^{-1} sr^{-1}$, are summarized in Fig. \ref{Fig3}. It is evident that at EeV in rock matter (as the one in Auger territory),  the expected rate exceed one event in three years. An enhancement, made by  peculiar Ande screen, may amplify the rate from the West side (doubling  the expected rate).
\begin{figure}
\begin{center}
\epsfig{figure=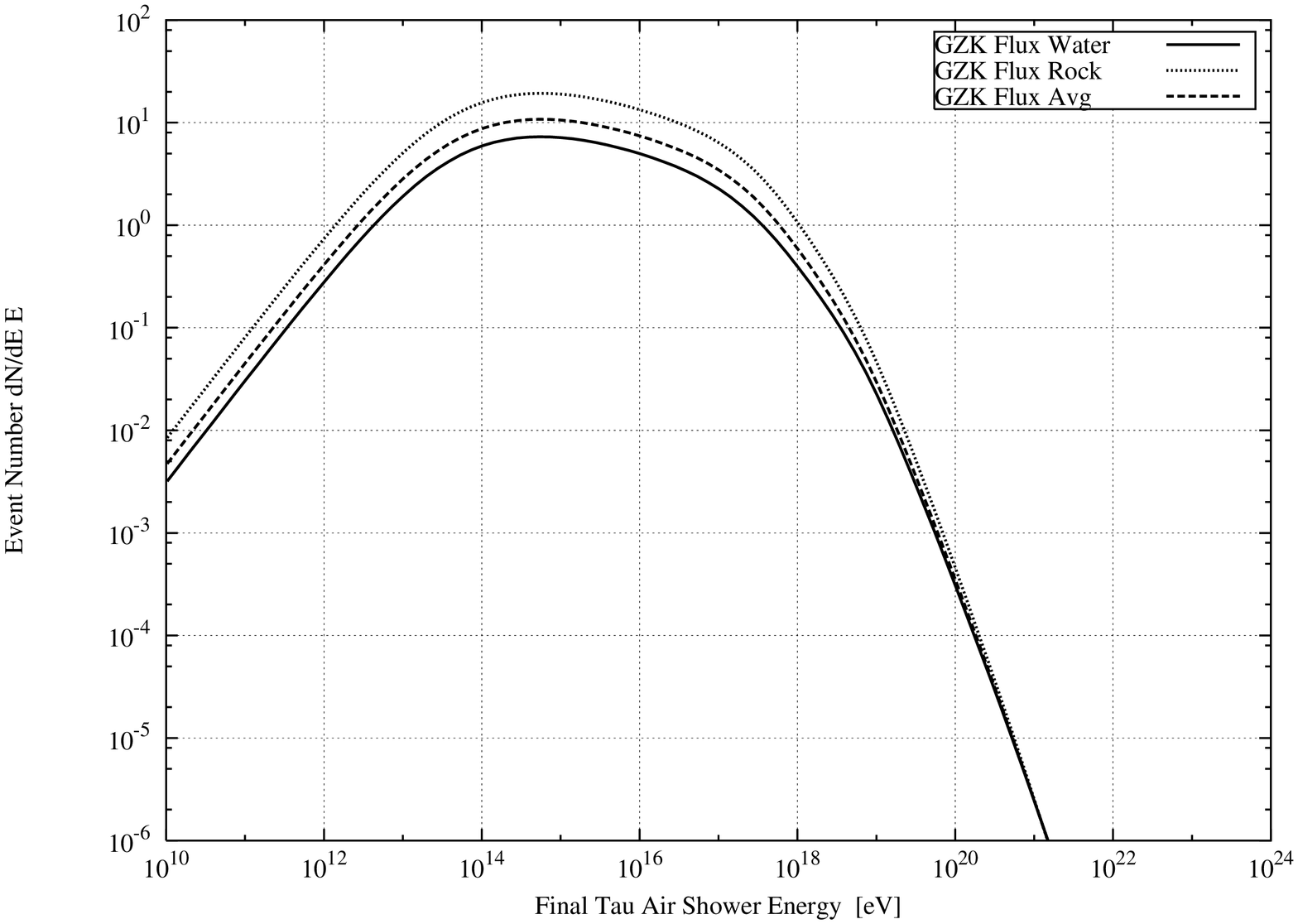,width=9.5cm,angle=-90}
\end{center}
\caption{ Our expected event rate spectra on Auger sky by
Fluoresce Detector in three years of records assuming an arrival
WB neutrino flux:$\Phi_{\nu_{\tau} +\overline{ \nu_{\tau}}}= 20 eV
\cdot cm^{-2}s^{-1} sr^{-1}$ \label{Fig3}. At EeV, where Auger FD the rate is
$N_{eV}= 1.07$ in three years; at $3 \cdot 10^{17}$  eV it reaches
$N_{eV}= 3.3$ ; at this energy the Auger acceptance is nearly a third of the area, reaching
once again the unity. It means that within present three years, i.e. this decade,
a Tau EeV event may rise in Auger sky within $ 2-0.3$ EeV . Additional event may occur
as inclined showers on surface detectors mostly arriving within  the Ande
shadow, a tau amplifier (double-triple rate from West than East side) observable by FD and SD.
Finally the extended horizontal and long tau air shower at high altitude (and low density) may be partially contained in Auger,increasing  the area and  the estimate above. Air shower Cherenkov  reflection on clouds may also be observable.}
\label{Fig3}
\begin{center}
\epsfig{figure=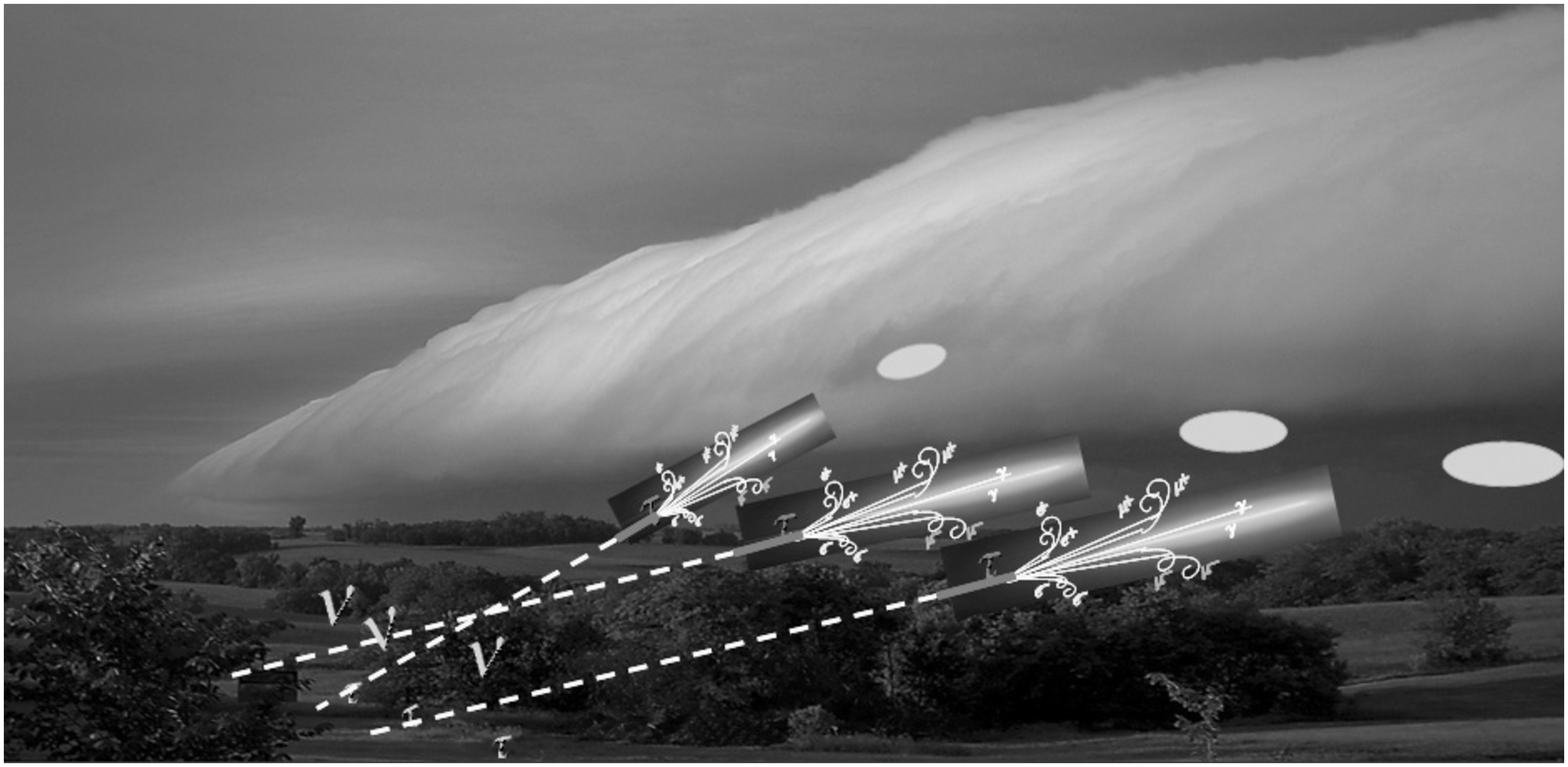,width=9.5cm}
\end{center}
\caption{ Upward Tau Air showering on the Auger clouds}
   \label{Fig4}
\end{figure}
  Inclined hadronic showers the more their zenith angle is large, the  higher
  their altitudes  take place. At highest quota (twenty-forty km), the air density
  is low, the pair threshold increases, the Cherenkov and Fluorescent luminosity
  decrees drastically. Moreover the distance from the high altitude till the Auger telescope
  increase and the hadronic high altitude Showers (Hias)\cite{Fargion1999} \cite{Fargion2005} are not longer observable by present array telescopes:  Auger is not able to reveal EeV hadronic air-shower above 75 degrees. Therefore, even within the poor Auger angular telescope  resolution any the inclined event within $80^o-100^o$, being long lived (because of the smaller air density ) air-showers must be indebt only to incoming UHE Neutrino. If upward, just tau ones.
\section{ Tau Air-Showers Rate: Young, in Ande Shadows, on Auger Sky}
  One of the most famous signature  of young Neutrino air-shower is their curvature
  and its time structure: it  may indicate the Tau  neutrino origin.\cite{Bertou2002}. However it is not the unique and most powerful imprint.  The Auger angular resolution and its limited statistics will not   allow to reveal any Moon or Solar Shadows. At least in a decade.
    The Ande shadow \cite{Fargion1999},\cite{Miele06} however is at least a thousand times larger than moon; however  on the horizons the UHECR rate decreases drastically, nearly three order of magnitude;  nevertheless the West-East asymmetry would rise around $88^o$ horizons as a few hundred  missing or asymmetric events, making meaningful its detection in one year.   It $must$ be observed soon by tuned trigger and angular resolution attention. Its  discover is an important crosscheck of the Auger  experiment.    Within this Ande shadow horizons taus might be better born, nearly one-two any three   year, mostly in FD (Fluorescence Detector), but also in SD (Surface Detector);  without any care on thresholds it will rise more rarely. In this decade Auger may find up-going Tau in its whole area  at the rate (see Fig. \ref{Fig3}) of $N_{10^{18}eV}=1.07$ event each three years ; at lower energy,  $N_{3\cdot 10^{17}eV}=  3 $  the Auger area detection is reduced ($\simeq 0.33$), leading to an important event rate  $N_{E_{\tau}= 3\cdot 10^{17}eV} \simeq 1.1$. Because  additional events  are un-confined (Horizontal) air-shower, this increases the detection mass  and its discover  rate, almost doubling the expectation rate. Moreover the presence  of an enhanced rate  from Ande size on FD and SD may increase the West side rate.   Finally a  possible  discover of FD could be amplified by  final flash  via Cherenkov reflection on clouds (see Fig. \ref{Fig4}). Being cloudy nights a third or a fourth of the whole time, this time may be an  occasion to exploit even if Moon arises. In conclusion, in partial disagreement to some earliest\cite{aramo05}  and most recent  Auger prospects \cite{Bigas07} requiring one or two $decades$ for a WB flux, we foresee,(in see also\cite{zas05}),  a sooner discover of GZK $\tau$ neutrino astronomy, possibly within  two-three years from now.  Auger may be even the first  experiment in the world to detect a tau  natural flavor regeneration processes. To reach and speed this goal we suggest: 1) to enlarge the telescope array facing towards the Ande. 2) To increase the array telescope angle of view, reducing the air-shower energy threshold, covering larger areas. 3) To tune the electronic trigger of FD to horizontal air-showers. 4) To map the UHECR Ande shadows at great angular resolution.

\end{document}